\documentclass[12pt]{article}
\usepackage{amsmath}
\usepackage{amsfonts}
\begin{document}

\vspace*{1cm}

\begin{center}
{\bf{\Large A Note on the Tachyon State in Vacuum String \\
\vspace*{.35cm}

Field Theory }}

\vspace*{1cm}

Radoslav Rashkov\footnote{e-mails: rrachkov@sfu.ca; rash@phys.uni-sofia.bg,
on leave of absence from Dept. of Physics, Sofia University, 1164 Sofia,
Bulgaria}
and K Sankaran Viswanathan\footnote{e-mail: kviswana@sfu.ca}\\

\ \\
Department of Physics, Simon Fraser University \\
Burnaby, BC, V5A 1S6, Canada
\end{center}

\vspace*{.8cm}

\begin{abstract}
We re-examine the recent proposal of Rastelli, Sen and Zwiebach on the tachyon
fluctuation of the vacuum string field theory representing a D25 brane,
originally considered by Hata and Kawano. We show that the tachyon state
satisfies the linearized equations of motion on-shell in the strong sense
thereby allowing us to calculate the ratio
$
\frac{
{\mathcal E}_c}
{{\mathcal T}_{25}}
$
of energy density to the tension of the D-brane to be
$
\frac{
{\mathcal E}_c}
{
{\mathcal T}_{25}}\simeq\frac{\pi^2}{3}\frac{1}{16(ln2)^3}
\simeq 0.62.
$
Our proof relies on a careful handling of the limits ($n\to\infty$) involved in
the conformal theory description of the sliver and tachyon states.
We conjecture that the sliver state represents a single D25 brane.
\end{abstract}

\vspace*{.8cm}

\section{Introduction}

In a recent paper \cite{hk} Hata and Kawano studied the fluctuation modes
around a classical solution to the vacuum string field theory (VSFT) called the
sliver. They identified a particular state (referred to hereafter as the HK
state) as an off-shell tachyon on the D25 brane solution and showed that
the linearized field equations around the classical background lead to the
correct on-shell condition for the tachyon. Furthermore, they calculated
the three tachyon amplitude using this on-shell state and found that the
energy density ${\mathcal E}_c$ comes to about roughly twice the D-brane
tension ${\mathcal T}_{25}$ \cite{hk}. They suggested that the sliver
perhaps
represents a state of two D-branes. In a very interesting paper \cite{rsz1}
Rastelli, Sen and Zwiebach reinvestigated the HK state using boundary
conformal field theory (BCFT) description as opposed to the Fock space
 description of Hata and Kawano. They propose a BCFT description of HK state
and present numerical evidence that this description agrees with the state
constructed in \cite{hk}. They further point out some problems in
the computations of the D25- brane tension. Specifically, they show that a
'naive' calculation of the D brane tension yields an answer that is twice
the expected value. However they show that the on-shell HK state fails to
satisfy the linearized equations of motion when one takes the inner product
with an HK state and hence point out that the 'naive' calculation is
incorrect.

In this work we reinvestigate the properties of the state discussed by
Rastelli, Sen
and Zwiebach. We show that one should pay particular care to the definition
of the BPZ product of the sliver state $|\chi_T(k)\rangle$ representing
a tachyon state with another sliver $|\chi_T(k')\rangle$. In particular
as these states are defined as $n\to\infty$ of wedge states in the CFT
language, care must be  exercised
in taking $n,n'\to\infty$ limits. We give a
consistent prescription for calculating both the star product of two slivers
and their BPZ product and show that the Hata-Kawano state (or more acurately,
Rastelli-Sen-Zwiebach's description in CFT language) satisfies the
linearized equations of motion in the strong sense as well as in weak sense.
More importantly, we calculate the ratio ${\mathcal E}_c/{\mathcal T}_{25}$
and find that it is given by
$\frac{{\mathcal E}_c}{{\mathcal T}_{25}}\simeq
\frac{\pi^2}{3}\frac{1}{16(ln2)^3}\simeq 0.62$.  Since this is a classical 
estimate, we conjecture that the sliver is a single brane solution.

\vspace*{.8cm}

\section{BCFT construction of the tachyon state}

The Vacuum String Field theory (VSFT) action is given by
\begin{equation}
S=-\kappa\left\{\frac 12\langle\Psi|Q|\Psi\rangle+\frac 13
\langle\Psi|\Psi\star\Psi\rangle\right\},
\label{1}
\end{equation}
where $|\Psi\rangle$ is the string field represented by a ghost number one
state in the matter-ghost BCFT, $Q$ is a new BRST operator of ghost number
one and made of ghost fields
\begin{equation}
Q=c_0+\sum\limits_{n\geq 1}f_n(c_n+(-1)^nc^\dagger_n).
\label{2}
\end{equation}
$\langle\Psi|\Phi\rangle$ represents the BPZ inner product and $\star$
denotes the star product \cite{w}.  The equations of motion are
\begin{equation}
Q|\Psi\rangle+|\Psi\star\Psi\rangle=0.
\label{3}
\end{equation}
Because of the special form of $Q$, one looks for a factorized solution
\begin{equation}
|\Psi\rangle=|\Psi_g\rangle\otimes|\Psi_m\rangle ,
\label{4}
\end{equation}
where $|\Psi_g\rangle$ denotes the ghost state and $|\Psi_m\rangle$ the
matter state. The equations of motion then read
\begin{align}
& Q|\Psi_g\rangle=|\Psi_g\star\Psi_g\rangle \notag \\
& |\Psi_m\rangle=|\Psi_m\star\Psi_m\rangle.
\label{5}
\end{align}
The ghost solution $|\Psi_g\rangle$ is taken to be universal and
$|\Psi_m\rangle$ obtained as a solution to the projector equation,
corresponds to different D-brane solutions. This interpretation follows from
the fact that with $Q$ constructed purely of ghost fields, it has trivial
cohomology and hence solutions to VSFT contain no perturbative open string
states, but may describe non-perturbative states such as the D-branes. In
several exciting papers \cite{rsz2,rsz3,rsz4} Rastelli, Sen and Zwiebach
discuss the properties and solutions to VSFT\footnote{For a recent nice review
see \cite{rsz5}}. They construct a solution to
the matter part of the equations of motion, describing a D25-brane solution,
called a sliver state $|\Xi_m\rangle$ defined through the relation
\cite{rsz2,rsz3}
\begin{equation}
\langle\Xi_m|\Phi\rangle= \lim_{n\to\infty}
\mathcal{N}
\langle f\circ\Phi(0)\rangle_{C_n},
\label{v5}
\end{equation}
where $f(z)=tan^{-1}z$, $|\Phi\rangle$ is an arbitrary state in the matter
Hilbert space, $\mathcal{N}$ is a normalization constant and
$\langle\dots\rangle_{C_n}$ denotes the correlation function of the matter
BCFT on a semi infinite cylinder $C_n$ of circumference $\frac{n\pi}{2}$
obtained
by making identification ${\mathcal Re}\,z\simeq
{\mathcal Re}\,z+\frac{n\pi}{2}$
in the upper half $z$ plane. In the $n\to\infty$ limit $C_n$ approaches the
upper half plane and
\begin{equation}
\langle\Xi_m|\Xi_m\rangle=KV^{(26)}, \qquad (V^{(26)}=(2\pi)^{26}
\delta^{26}(0))
\label{7}
\end{equation}
where $V^{(26)}$ is the volume of 26-dimensional spacetime and $K$ is a
normalization constant that arises due to anomaly in the matter sector.

Rastelli, Sen and Zwiebach propose that the HK tachyon state can be written
in the form
$$
|\Psi_g\rangle\otimes|\chi_T(k)\rangle
$$
where $|\Psi_g\rangle$ is the same state as in eq. (\ref{4}) and
$|\chi_T(k)\rangle$ is the matter part defined through the
relation
\begin{equation}
\langle\chi_T(k)|\Psi\rangle=\mathcal{N}\lim_{n\to\infty} n^{2k^2}
\langle e^{ik.X(\frac{n\pi}{4})}f\circ\Psi(0)\rangle_{C_n} .
\label{def}
\end{equation}
Here  $|\Psi\rangle$ is any state in the Hilbert space of states of the
string field.
This relation tells us to insert a tachyon vertex in the middle of the
sliver at $n\pi/4$ which is diametrically opposite the puncture at the origin.
BPZ and star products of two sliver states can be constructed by cutting
and pasting of the cylinders $C_n$ and $C_{n'}$ according to the
prescription in \cite{rsz2,rsz5}.

For arbitrary $|\Psi\rangle$ (\ref{def}) can be calculated following
\cite{rsz1} by writting
\begin{equation}
f\circ\Psi(0)=a_\Psi e^{-ik.X(0)}+
\left[descendents\,\, of\,\, e^{-ik.X(0)}\right] + \cdots
\label{10}
\end{equation}
 Only the first term carrying momentum $-k$ contributes to (\ref{def})
 in the $n\to\infty$ limit as $e^{-ik.X(0)}$ is the only primary operator
 non orthogonal to the insertion $e^{ik.X}$.
One finds readily upon evaluating the correlation function on right side of
(\ref{def}) that
\begin{equation}
\langle\chi_T(k)|\Psi\rangle=\mathcal{N} 2^{2k^2}a_\Psi V^{(26)}
\label{def1}
\end{equation}
by using the procedure for evaluating star and BPZ products of two sliver
states.
 It is established in \cite{rsz1} that the state $|\chi_T(k)\rangle$ satisfies
the linearized equations of motion on-shell (i.e. $k^2 =1$) in the weak form
\begin{equation}
\langle\chi_T(k)|\Psi\rangle=
\langle\Xi_m\star\chi_T(k)+\chi_T(k)\star\Xi_m|\Psi\rangle .
\label{v9}
\end{equation}
However they show that eq. (\ref{v9}) ceases
to be valid if the Hilbert space
state $|\Psi\rangle$ is replaced by another sliver
$|\chi_T(k')\rangle$ (i.e. Fock space state). Since this product arises
in the action for the tachyon they conclude that the result for the ratio
${\mathcal E}_c/{\mathcal T}_{25}$ is incorrect.
We now show that the linearized equations in the form
\begin{equation}
\langle\chi_T(k)|\chi_T(k')\rangle=
\langle\Xi\star\chi_T(k)+\chi_T(k)\star\Xi|\chi_T(k')\rangle
\label{v10}
\end{equation}
is actually valid on-shell if we use proper caution in taking
$n,n'\to\infty$ limits.
According to the rules in \cite{rsz2,rsz5}
\begin{equation}
\langle\chi_T(k)|\chi_T(k')\rangle=
\lim_{n\to\infty, n'\to\infty} n^{2k^2}{n'}^{2{k'}^2}
\langle e^{ik'.X\left(
(n+n'-2)\frac{\pi}{4}\right)} e^{ik.X(0)}\rangle_{C_{n+n'-2}} .
\label{v11}
\end{equation}
At first glance the definition of BPZ product in (\ref{def}) and
(\ref{v11}) appear to involve different rules. But careful look shows that
they are in fact consistent. Let us demonstrate the equivalence of
(\ref{v11}) with (\ref{def}). The normalization constants in (\ref{v11})
are of the form $(\frac{length}{\pi/2})^{2k^2}$ where the length refers to
the circumference of the cylinders defining the states $|\Psi\rangle$ as
surface states in a boundary conformal field theory. In (\ref{v11}) the
two cylinders defining $|\chi_T(k)\rangle$ and $|\chi_T(k')\rangle$ have
circumferences $(n-1)\frac{\pi}{2}$ and $(n'-1)\frac{\pi}{2}$ respectively.
Hence take for the moment these normalization factors to be $l_n=n-1$ and
$l_{n'}=n'-1$. We now make a change of variables to $\tilde z=z/(n'-1)$.
Then the length of the first strip representing $|\chi_T(k')\rangle$ in
the BPZ product becomes $\pi/2$ while the length of the second strip
is $(\tilde n-\varepsilon_{n'})\frac{\pi}{2}$, where $\tilde n=n/(n'-1)$
and $\varepsilon_{n'}=1/(n'-1)$. Then, a simple computation yields
\begin{equation}
\langle\chi_T(k)|\chi_T(k')\rangle=
Kl_{n'}^{2{k'}^2}\tilde l_{\tilde n}^{2k^2}
\left(\frac{1}{n'-1}\right)^{k^2+{k'}^2}
\langle e^{ik'.X\left((\tilde n+1-\varepsilon_{n'})\frac{\pi}{4}\right)}
e^{ik.X(0)}\rangle_{C_{\tilde n+1-\varepsilon_{n'}}} .
\label{norm}
\end{equation}
Now in the limit $\tilde n, n'\to\infty$ this become
\begin{equation}
\langle\chi_T(k)|\chi_T(k')\rangle=
\lim_{n\to\infty} n^{2k^2}
\langle e^{ik'.X(n\frac{\pi}{4})} e^{ik.X(0)}\rangle_{C_n}
\label{norm1}
\end{equation}
which is a special case of (\ref{v11}). The above correlation function can
be evaluated after mapping $C_n$ into unit disk $D$ and we find
\begin{equation}
\langle\chi_T(k)|\chi_T(k')\rangle
=K2^{2k^2}(2\pi)^{26}\delta(k+k') .
\label{v13}
\end{equation}

In evaluating $\langle\Xi_m\star\chi_T(k)+
\chi_T(k)\star\Xi_m |\chi_T(k')\rangle$
we first express $\langle\Xi_m\star\chi_T(k)\rangle$ (and
$|\chi_T(k)\star\Xi_m\rangle$) as a sliver (i.e we let $n_2=n_3=n$ in these
states and $\tilde n$ as in the above) we then find
\begin{align}
\langle\Xi_m &\star\chi_T(k)|\chi_T(k')\rangle \notag \\
&=K\lim_{n_1\to\infty}n_1^{2{k'}^2}\left[
\lim_{n_2,n_3\to\infty}n_2^{2k^2}
\langle e^{ik.X\left((2n_2+n_3+n'-4)\frac{\pi}{4}\right)}
e^{ik'.X(0)}\rangle_{C_{n'+n_2+n_3-3}}
\right] \notag \\
&=K\lim_{\tilde n\to\infty}\tilde n^{2k^2}
\langle e^{ik.X\left((3\tilde n+1-3\varepsilon_{n_1})\frac{\pi}{4}\right)}
e^{ik'.X(0)}\rangle_{C_{2\tilde n+1-2\varepsilon_{n_1}}} .
\label{v14}
\end{align}
We change to a new variable
$w=exp(\frac{4iz}{2\tilde n+1-2\varepsilon_{n_1}})$ so that
$C_{2\tilde n+1-2\varepsilon_{n_1}}$ is mapped
to a unit disk $D$. We let $\tilde n\to\infty$  and take also $n_1\to\infty$.
We obtain
\begin{align}
\langle\Xi_m&\star\chi_T(k)|\chi_T(k')\rangle  \notag \\
=& K\lim_{n_1\to\infty}
\lim_{\tilde n\to\infty}\left\{
n^{2k^2}\left(\frac{4}{2\tilde n+1-2\varepsilon_{n_1}}\right)^{k^2+{k'}^2}
\langle e^{ik.X\left(e^{\frac{3i\pi}{2}}\right)}
e^{ik'.X(1)}\rangle_D
\right\} \notag \\
=& K\,2^{k^2}(2\pi)^{26}\delta(k+k') .
\label{v15}
\end{align}
$\langle\chi_T(k)\star\Xi_m|\chi_T(k')\rangle$ is calculated similarly and
is equal to (\ref{v15}). We find
\begin{equation}
\langle\Xi_m\star\chi_T(k)+\chi_T(k)\star\Xi_m|\chi_T(k')\rangle
= K\, 2^{k^2+1}(2\pi)^{26}\delta(k+k').
\label{vv15}
\end{equation}
Comparing (\ref{vv15}) with equation (\ref{v13}) we see that
\begin{equation}
2^{1-k^2}\langle\chi_T(k)|\chi_T(k')\rangle=
\langle\Xi_m\star\chi_T(k)+\chi_T(k)\star\Xi_m
|\chi_T(k')\rangle
\label{v16}
\end{equation}
which for $k^2=1$ reduces to eq. (\ref{v9}), thus giving
the correct on-shell condition for the tachyon on the D-brane.
Thus the tachyon state $|\chi_T(k)\rangle$ satisfies the linearized equations
of motion in the strong sense. As we notice from the above steps, it
is crucial to take limits $n,n'\to\infty$ separately and not let
$n=n'\to\infty$. The above calculations lead us to the following formal rules
in manipulating with sliver states:

a) When we take a star product of two slivers, it is permissible to
let $n_1=n_2=n\to\infty$ and express the starproduct state as a sliver.

b) When we take the BPZ product of two slivers one of the
limits $n\to\infty$ (keeping the other $n'$ large but fixed) first and
express the result as a limit over the other $n'$.

These rules  correspond to the procedure explained above and used in 
calculating (\ref{v13}) and (\ref{vv15}).

\vspace*{.8cm}

\section{The D25-brane tension and  energy density}

In this section we compute the D25-brane tension by using its relation
to the three tachyon coupling $g_T$ \cite{s,gs}.

If the classical solution $|\Psi_m\rangle$ to VSFT were to represent a
D25-brane, then the ratio of the energy density ${\mathcal E}_c$ to
the brane tension $\mathcal T_{25}$ is expected to be unity. Previous
estimates
\cite{rsz1} and \cite{hk,hm} give roughly 2, giving rise to the speculation
that
$|\Xi_m\rangle$ in fact is a solution representing two D25 branes.

To compute the tachyon coupling we must determine the quadratic and cubic
terms in the tachyon field arising in the action (\ref{1}).

We follow \cite{rsz1} in using the general expansion
\begin{equation}
|\Psi\rangle=|\Psi_g\rangle\otimes\left\{
|\Xi_m\rangle +\int d^{26}k \, n^{-k^2}T(k)|\chi_T(k)\rangle+\cdots
\right\} .
\label{18}
\end{equation}
Here $T(k)$ is the tachyonic field amplitude and ellipses denote of
higher order excitations. Subsituting (\ref{18}) into (\ref{1}) we get
\cite{rsz1}
\begin{align}
S=& S(|\Psi_g\rangle\otimes|\Xi_m\rangle)-\langle\Psi_g|Q|\Psi\rangle
\times \notag \\
& \Bigl\{
\frac 12\int d^{26}k\, d^{26}{k'}T(k)T(k')
\langle\chi_T(k')|
\bigl[|\chi_T(k)\rangle
-|\Xi_m\star\chi_T(k)\rangle
-|\chi_T(k)\star\Xi_m\rangle
\bigr]
  \notag \\
& +\frac 13\int d^{26} k_1\,d^{26} k_2\,d^{26}\, k_3T(k_1)T(k_2)T(k_3)
\langle\chi_T(k_1)|\chi_T(k_2)\star\chi_T(k_3)\rangle
\Bigr\} .
\label{19}
\end{align}
From equations (\ref{v13}) and (\ref{vv15}) we have for off-shell the relation
\begin{equation}
2^{1-k^2}|\chi_T(k_3)\rangle=|\Xi\star\chi_T(k)\rangle+ 
|\chi_T(k)\star\Xi\rangle .
\label{20}
\end{equation}
The quadratic term in the action can now be written as
\begin{align}
S^{(2)}=&-\frac 12\langle\Psi_g|Q|\Psi\rangle
\int d^{26}k \, d^{26}k' \, T(k)T(k')
\left[1-2^{1-k^2}\right]
\langle\chi_T(k')|\chi_T(k)\rangle
 \notag \\
\simeq &\frac 12\langle\Psi_g|Q|\Psi\rangle\, ln2
\int d^{26}k\, d^{26}k' \, T(k)T(k')[k^2-1]
\langle\chi_T(k')|\chi_T(k)\rangle ,
\label{21}
\end{align}
where in the second step we have taken the tachyon field $T(k)$
near on-shell $k^2\simeq 1$.

Subsituting for the inner product from (\ref{v13}) we find for the quadratic
term
\begin{equation}
S^{(2)}\simeq\frac K2\, 4ln2\,
\langle\Psi_g|Q|\Psi\rangle (2\pi)^{26}
\int d^{26}k\, d^{26}k' \, T(-k)T(k)[k^2-1] .
\label{22}
\end{equation}
By a redefinition of $T(k)$
\begin{equation}
\hat T(k)=2\sqrt{K\,ln2\,\langle\Psi_g|Q|\Psi_g\rangle} T(k)
\label{24}
\end{equation}
we can write $S^{(2)}$ as
\begin{equation}
S^{(2)}\simeq\frac 12(2\pi)^{26}\int d^{26}k\, (k^2-1)\hat T(-k)\hat T(k) .
\label{23}
\end{equation}
Consider next the cubic term
\begin{equation}
S^{(3)}=
\frac 13\int d^{26}k_1\, d^{26}k_2\, d^{26}k_3 \, T(k_1)T(k_2)T(k_3)
\langle\chi_T(k_1)|\chi_T(k_2)\star\chi_T(k_3)\rangle .
\label{25}
\end{equation}
The BPZ product in (\ref{25}) can be readily calculated following
\cite{rsz1} and by using the rules outlined in the previous section.
\begin{align}
&\langle\chi_T(k_1)|\chi_T(k_2)\star\chi_T(k_3)\rangle=
\notag \\
&= Kn_1^{2k_1^2}n_2^{2k_2^2}n_3^{2k_3^2}\langle e^{ik_1.X(0)}
e^{ik_2.X\left((\tilde n+1-\varepsilon_{n_1})\frac{\pi}{4}\right)}
e^{ik_2.X\left((3\,\tilde n+1-3\varepsilon_{n_1})\frac{\pi}{4}\right)}
\rangle_{C_{2\tilde n +1-2\varepsilon_{n_1}}}
 \notag \\
&=K\,\tilde n^{2k_2^2+2k_3^2}
\left(\frac{4}{2\,\tilde n+1-2\varepsilon_{n_1}}\right)^{k_1^2+k_2^2+k_3^2}
\langle e^{k_1.X(1)}e^{ik_2.X(e^{i\pi/2})}e^{ik_2.X(e^{i3\pi/2})}
\rangle_D ,
\label{26}
\end{align}
where in the last step we  used a change of variable
$w=e^{i4z/(2\,\tilde n+\varepsilon_{n_1}-3)}$ to map the cylinder
$C_{2\tilde n +\varepsilon_{n_1}-3}$
into a unit disk $D$. Evaluating the correlation function
on $D$ we find that
\begin{align}
\langle\chi_T(k_1)|\chi_T(k_2)\star\chi_T(k_3)\rangle
\simeq & \tilde n^{k_2^2+k_3^2-k_1^2}2^{k_1^2}
(2\pi)^{26}\delta(k_1+k_2+k_3)
 \notag \\
\simeq & \tilde n^{k_2^2+k_3^2-k_1^2}2
(2\pi)^{26}\delta(k_1+k_2+k_3) .
\label{27}
\end{align}
Thus the cubic term in the action near on-shell becomes
\begin{align}
S^{(3)}\simeq&
-2\frac K3(2\pi)^{26}\langle\Psi_g|Q|\Psi_g\rangle
\int d^{26}k_1\, d^{26}k_2\, d^{26}k_3\, \delta(k_1+k_2+k_3)
\, T(k_1)T(k_2)T(k_3) \notag \\
=&-\frac 13\frac{2(2\pi)^{26}}{(2ln2)^3\sqrt{K\langle\Psi_g|Q|\Psi_g\rangle}}
\int d^{26}k_1\, d^{26}k_2\, d^{26}k_3\, \hat T(k_1)\hat T(k_2)\hat T(k_3)
\,\delta(k_1+k_2+k_3).
\label{28}
\end{align}
The on-shell three tachyon coupling is therefore given by
\begin{equation}
g_T=\frac{2}{(2\sqrt{ln2})^3}\frac{1}
{\sqrt{K\langle\Psi_g|Q|\Psi_g\rangle}}\, .
\label{30}
\end{equation}
According to \cite{s,gs} the D-brane tension is related to the above coupling
$g_T$ in the following way
\begin{equation}
{\mathcal T_{25}}=\frac{1}{2\pi^2 g_T^2}=
\frac{1}{2\pi^2}K\langle\Psi_g|Q|\Psi_g\rangle
\frac{(2\sqrt{ln2})^6}{4}
\label{31}
\end{equation}
we note that our expression for the brane tension ${\mathcal T}_{25}$
differs
from that in \cite{rsz1}. We recall also the expression for the energy density
${\mathcal E}_c$ corresponding to the sliver
\begin{equation}
{\mathcal E}_c=\frac K6\langle\Psi_g|Q|\Psi_g\rangle .
\label{32}
\end{equation}
From (\ref{31}) and (\ref{32}) we get for the ratio of the energy density
 to the brane tension
\begin{equation}
\frac{
{\mathcal E}_c}
{
{\mathcal T}_{25}} =\frac{\pi^2}{3}\frac{1}{16(ln2)^3}
\simeq 0.62.
\label{33}
\end{equation}

\vspace*{.8cm}

\section{Conclusions}

In view of the fact that the calculated value of
$\mathcal{E}_c/\mathcal{T}_{25}$
is less than unity we are inclined to conjecture that the sliver state
does in fact represent a single D25 brane. We can attribute the deviation
from unity to the fact that the result (\ref{33})
is a perturbative estimate.

\vspace*{.8cm}

{\bf Acknowledgements:}  R.R. would like to thank Simon Fraser
University for warm hospitality. This work has been supported by an
operating grant from the Natural Sciences and Engineering Research Council
of Canada.

\end{document}